\documentclass[twocolumn,twoside,slac_two]{revtex4}
\usepackage{graphicx}
\usepackage{fancyhdr}
\usepackage{multirow}
\usepackage[tight]{subfigure}
\usepackage{color}
\usepackage{array}

\begin{document}

\title{The VERITAS Survey of the Cygnus Region of the Galactic Plane}

%

\author{Amanda Weinstein for the VERITAS Collaboration}
\affiliation{University of California Los Angeles, 475 Portola
Plaza, Los Angeles, CA 90095 USA }
\begin{abstract}
The VERITAS IACT observatory has carried out an extensive survey of
the Cygnus region between 67 and 82 degrees in galactic longitude
and between -1 and 4 degrees in galactic latitude. This region is a
natural choice for a Very High Energy (VHE) gamma-ray survey in the
Northern Hemisphere, as it contains a substantial number of
potential VHE gamma-ray emitters such as supernova remnants, pulsar
wind nebulae, high-mass X-ray binaries, and massive star clusters,
in addition to a few previously detected VHE gamma-ray sources. It
is also home to a number of GeV gamma-ray sources, including no less
than four new high-significance sources detected in the first six
months of Fermi data. The VERITAS survey, comprising more than 140
hours of observations, reaches an average VHE point-source flux
sensitivity of better than 4\% of the Crab Nebula flux at energies
above 200 GeV. Here we report on preliminary results from this
survey, including two source detections, and discuss the prospects
for further studies that would exploit the joint coverage provided
by VERITAS and Fermi data in this region.

\end{abstract}

\maketitle

\thispagestyle{fancy}


\section{Introduction}

To date, only a few moderate-scale surveys have been performed in
gamma rays between $\rm 100 \thinspace GeV$ and $\rm 10 \thinspace
TeV$: the HESS scan of the central region of the Galactic
Plane\cite{hess_survey}, the much less sensitive HEGRA survey of the
quarter of the Galactic Plane between $-2^{\circ}$ and $85^{\circ}$
in galactic longitude\cite{hegra}, and the VERITAS scan of the
Cygnus region under discussion here.  However, there is a strong
motivation for performing such surveys; an unbiased search of a
substantial region of sky is less subject to the experimental and
theoretical prejudices that guide most VHE gamma-ray observations,
and therefore (as demonstrated by HESS) offers greater scope for
serendipitous discoveries.  An unbiased search also allows for more
quantitative statements to be made about the source population in
the region surveyed.

The stereoscopic imaging atmospheric Cherenkov telescope (IACT)
array VERITAS has just completed a two-year survey of a 5 by 15
degree portion of the Cygnus region of the Galactic Plane.  The
Cygnus region was a natural target for survey observations, as it is
already known to contain a significant number of potential TeV
gamma-ray emitters.  In the GeV ($\rm 20 MeV-300$ GeV) energy band,
it is home to a number of sources or potential sources, including no
less than 4 distinct Fermi sources \cite{fermibrightsource}.
Moreover, both Fermi and its predecessor EGRET have detected diffuse
emission from this region that is greater in flux than all of the
currently resolved sources taken together. Viewed in the energy
range between 1TeV-50 TeV, it contains a pair of unidentified TeV
sources (MGRO J2031+41 and MGRO 2019+37) detected by the Milagro
Gamma Ray Observatory, a water Cherenkov extensive air shower array
that is sensitive to TeV sources at a median energy of 20 TeV over
the entire sky \cite{MGROsurvey}, as well as the unidentified source
TeV J2032+4130 (first detected by the HEGRA IACT
array\cite{HEGRAtev}), that is spatially coincident with MGRO
J2031+41.  The exact nature of these sources is currently unknown.
There is also a significant catalog of objects detected at other
wavelengths, including SNRs, pulsar wind nebulae (PWNe), high-mass
x-ray binaries (HMXBs) and massive star clusters, that are
considered potential TeV sources.

\section{Survey Observations}

\begin{figure*}[!t]
  \centering
  \includegraphics[width=5in]{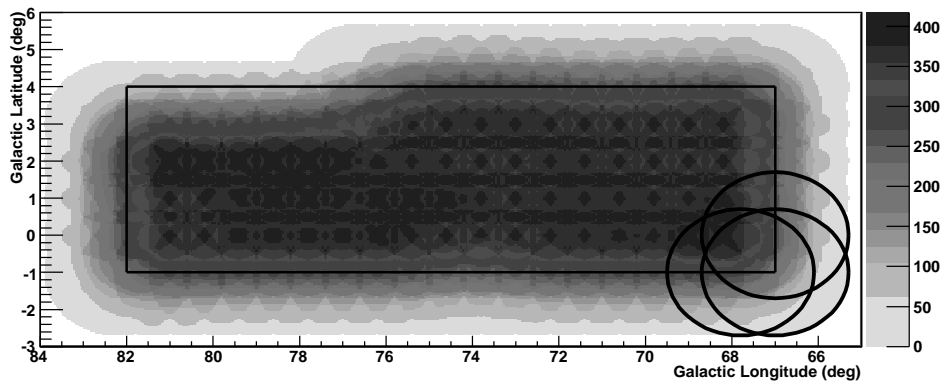}
  \caption{Effective exposure map for the VERITAS Cygnus sky survey before
  follow-up observations, based on data obtained from Spring 2007 through Fall 2008. The color
scale to the right gives the effective exposure time in minutes. The
black box indicates the boundary of the survey region proper; the
black circles show examples of the overlapping fields of observation
that are tiled to produce the survey.}
  \label{survey_exposure}
 \end{figure*}

Survey observations, which began in April 2007 and were completed in
November 2008, cover the field between galactic longitudes
$67\,^{\circ}$ and $82\,^{\circ}$ and galactic latitudes
$-1\,^{\circ}$ and $4\,^{\circ}$ with a grid of pointed
observations.  Grid points have $0.8\,^{\circ}$ separation in
galactic longitude and $1\,^{\circ}$ separation in galactic
latitude, allowing for substantial overlap in the fields of view for
observations at nearby grid points.  Approximately one hour of
observing time was taken at every grid point (generally within a 1-3
day period), with that hour broken into 20-minute observation
periods. Figure \ref{survey_exposure} shows both a schematic of the
survey observation strategy and the acceptance-corrected
(``effective'') exposure time for the ``base'' survey (that is, the
survey data taken prior to follow-up observations) over the entire
survey field.  The base survey achieves a relatively uniform
effective exposure of $\sim 6-7$ hours.  The full survey dataset
($>140$ hours of good-quality observation time) has regions of
enhanced exposure due to follow-up observations taken in Fall 2008,
Spring 2009, and Fall 2009.

Survey data have been quality-selected to remove runs with unstable trigger
rates, poor weather conditions, and known hardware problems.  Almost all survey
data were taken on moon-less nights, with a few percent taken under slight to
moderate moonlight conditions.  Since survey observations began during the
commissioning period for the VERITAS array, data in the base survey were taken
in two different configurations.  Observations from Spring 2007 were taken with
a three-telescope array configuration (including telescopes 1, 2, and 3) and
observations from Fall 2007 and Spring and Fall 2008 were taken with the full
four-telescope array.

Survey observations were taken over a range of zenith angles between
$10^{\circ}$ to $35^{\circ}.$  Observations were scheduled in such a way as to
keep the average zenith angle of observations for any survey pointing at
$20^{\circ}$, but constraints in terms of available time did not always allow
for this.  We estimate that $45\%$ of the survey region is covered by
observations taken at an average zenith angle of $20^{\circ}$ or smaller,
another $33.5\%$ by observations taken at $20^{\circ}  < z < 25^{\circ}$, $9\%$
by observations taken at $25^{\circ} < z <30^{\circ}$, and $12.5\%$ by
observations taken at $z > 30^{\circ}$.

\subsection{Follow-up Observations}

Follow-up data taken in Fall 2008 were also taken with the four-telescope array
configuration: observations during this period are a combination of two types
of observations: wobble observations at $0.5^{\circ}$ to $0.7^{\circ}$ offsets
from points of interest ($>4\sigma$ fluctuations or locations of reported AGILE
or INTEGRAL transients) within the survey region, and additional exposure on
survey pointings near a point or region of interest.  Further follow-up
observations ($~60-70$ hours total) using a wobble strategy were taken in
Spring and Fall 2009.  Due to a scheduled VERITAS upgrade, data taken in Spring
2009 were taken with a three-telescope array configuration containing
telescopes 2, 3, and 4, while data taken in Fall 2009 was taken with the full
array in an upgraded configuration, where telescope 1 has been moved to improve
the instrument baselines.  This upgrade has improved the sensitivity of the
array to weak point sources by about $20\%$\cite{vsensitivity}.

\section{Survey Analysis}

Given that the H.E.S.S. survey revealed a population of Galactic TeV
gamma-ray emitters that was biased towards hard-spectrum, moderately
extended sources\cite{hess_survey}, and the fact that that the
Milagro survey of the Cygnus region \cite{MGROsurvey} showed sources
of significant apparent extent, it was reasonable to expect that
some or all of the visible gamma-ray sources in this region would
also have relatively hard spectra and significant extension.

In order to optimize the survey's sensitivity to such sources while
not sacrificing too much in the way of sensitivity to
softer-spectrum and/or point-like sources, a set of parallel
analyses was used.  Each analysis is a variation on a common base
analysis, optimized for better sensitivity to a particular type of
source. In order to limit the number of additional trials factors
incurred, the number of parallel analyses was restricted to four.
Two were optimized for sources with a Crab-like spectrum---one for
point sources and the other for moderately extended
($r=0.2^{\circ}$) sources---and two more were optimized for
hard-spectrum point-like and extended sources, using a reference
source with a power-law spectrum and a spectral index of $2.0$.

The completely common elements of the analysis procedure consist of
calibrating and cleaning the Cherenkov images and parameterizing
them by second moments\cite{hillas}.  The technique used to
stereoscopically reconstruct the shower direction and impact
parameter is likewise common to all analyses: however, images used
in stereoscopic reconstruction are required to exceed a minimum
integrated charge (\emph{size}) in digital counts (dc), and the
value of that requirement is analysis-specific as shown in Table
\ref{cut_summary}.  Showers are reconstructed for all events in
which at least three telescope images passed quality selection.


 \begin{table}[!h]
  \caption{Cuts Specific to Parallel Analyses}
  \label{cut_summary}
  \centering
  \begin{tabular}{|c|p{2.1cm}|p{2.0cm}|}
  \hline
   \hfil  &  Point Source & Extended Source \\ \hline
   \multirow{2}{1.8cm}{Standard Source}
    & $\rm size > 600 dc$ \quad\quad $\rm (\sim 90 \thinspace p.e.)$  & $\rm size > 600 dc$ \\
    & $\theta^2 < 0.013$ &   $\theta^2 < 0.055$
    \\ \hline
   \multirow{2}{1.8cm}{Hard Source}
    & $\rm size > 1000 dc$ \quad\quad$\rm (\sim 150 p.e.)$  & $\rm size > 1000 dc$ \\
    & $\theta^2 < 0.013$ &   $\theta^2 < 0.055$
    \\ \hline

  \end{tabular}
  \end{table}

Cosmic ray background is rejected using two means.  The first is a
pair of variables (\emph{mean scaled length} (MSL) and \emph{mean
scaled width} (MSW) that summarize differences in image shape
between gamma ray events and the majority of the cosmic ray
background)\cite{mslw} and are applied prior to generation of photon
sky maps.  The second is a cut on the square of the angular distance
($\theta^2$) between a reconstructed shower and the sky position of
a potential source that is applied as part of that process.  The
cuts on \emph{mean scaled length} and \emph{mean scaled width} are
common to all parallel analyses ($\rm 0.05 < MSW < 1.06$, $\rm 0.05
< MSL < 1.24$), while the cut in $\theta^2$ is analysis-specific as
shown in Table \ref{cut_summary}. The residual cosmic ray background
is estimated using the ``ring-background'' model~\cite{ring}.

\section{Survey Detections}

\begin{figure*}[!t]
  \centering
  \includegraphics[width=4in, trim=0 0 0 4]{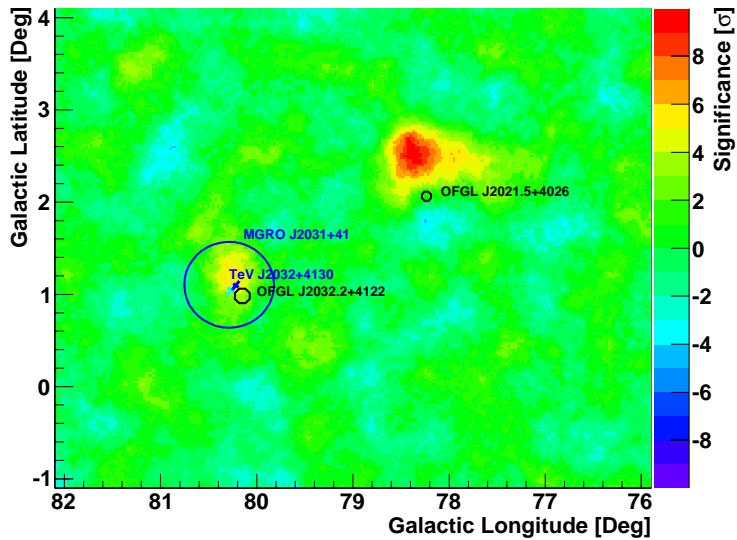}
  \caption{Significance map of the portion of the VERITAS Cygnus Region
  Sky above $\rm 76^{\circ}$ galactic longitude, including all follow-up data taken through Fall 2009.  The
  error circles of two sources from the Fermi six-month bright
  source list\cite{fermibrightsource}, OFGL J2021.5+4026 and OFGL J2032.2+4122, both pulsars,
  are indicated by black circles, and the position and extent of
  MGRO J2031+41\cite{MGROsurvey} is indicated in blue.  The position of TeV
  J2032+4130 as measured by HEGRA and MAGIC is indicated by the dark
  and light blue crosses, respectively\cite{HEGRAtev}\cite{MAGICtev}.
  }
  \label{survey_map}
 \end{figure*}

Figure \ref{survey_map} shows the pre-trials significance at all
points in the VERITAS Cygnus Region survey above $\rm 76^{\circ}$
galactic longitude, based on the standard extended source analysis
and including all data taken through Fall 2009.  This region of the
survey contains two clear detections of VHE gamma-ray emission: the
known source TeV J2032+4130\cite{HEGRAtev}, and a newly discovered
source, VER J2019+407, in the vicinity of the $\gamma$ Cygni SNR. As
shown in Figure \ref{survey_map}, both sources are in proximity to a
Fermi pulsar, and TeV J2032+4130 is also coincident with the
multi-TeV gamma-ray source MGRO 2031+41.

In the full survey dataset, VER J2019+407 is detected at $10\sigma$ pre-trials
and TeV J203+4130 is detected at $5\sigma$ at the nominal position.  Neither
source exceeded $\rm 5\sigma$ pre-trials in the base survey.

\subsection{VER J2019+407}

\begin{figure*}[ht]
  \centering \subfigure[] {%
  \includegraphics[width=3in, trim=0 0 0 4]{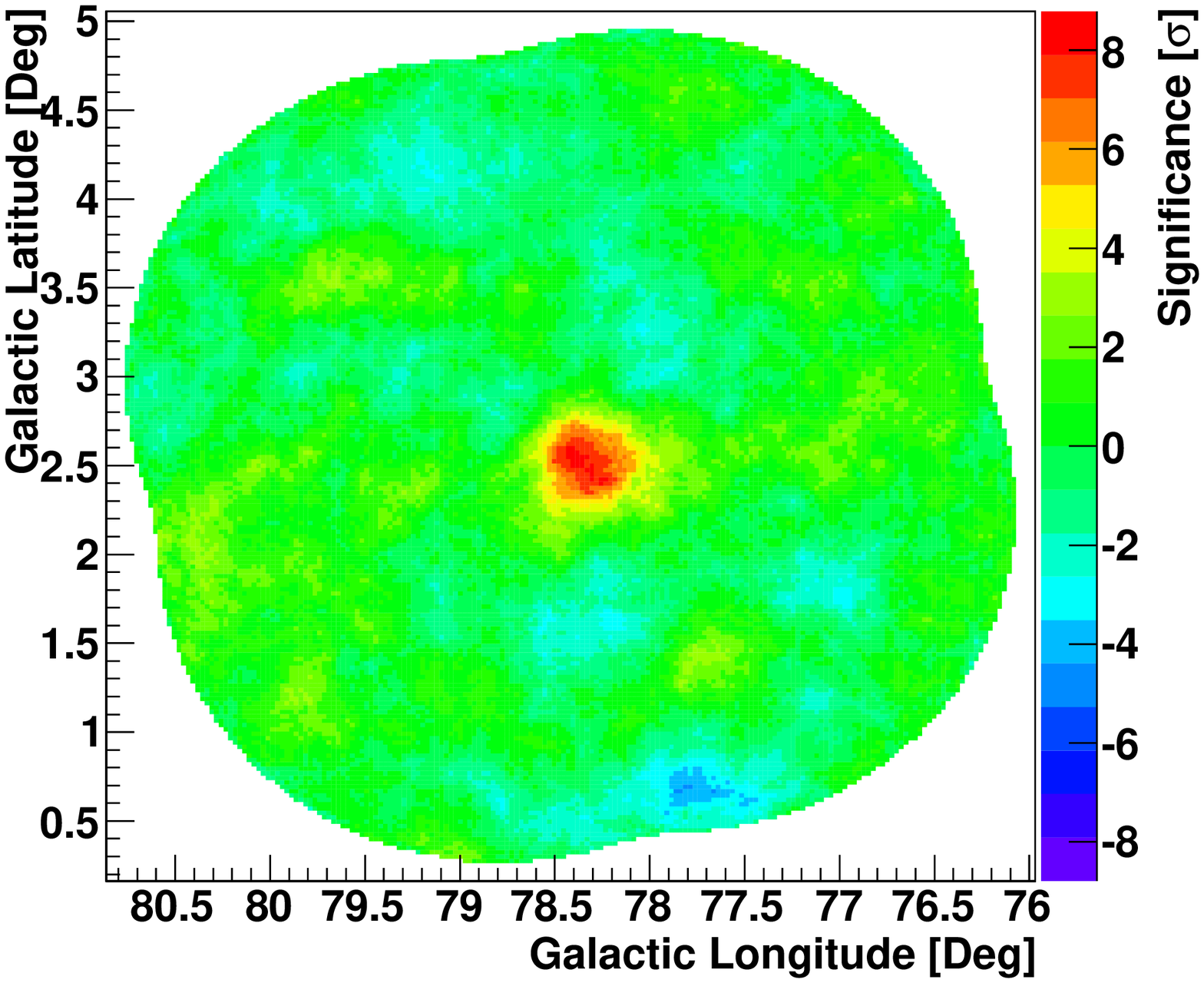}
  \label{cygni_map}}
  \subfigure[] {%
  \includegraphics[width=3in, trim=0 0 0 4]{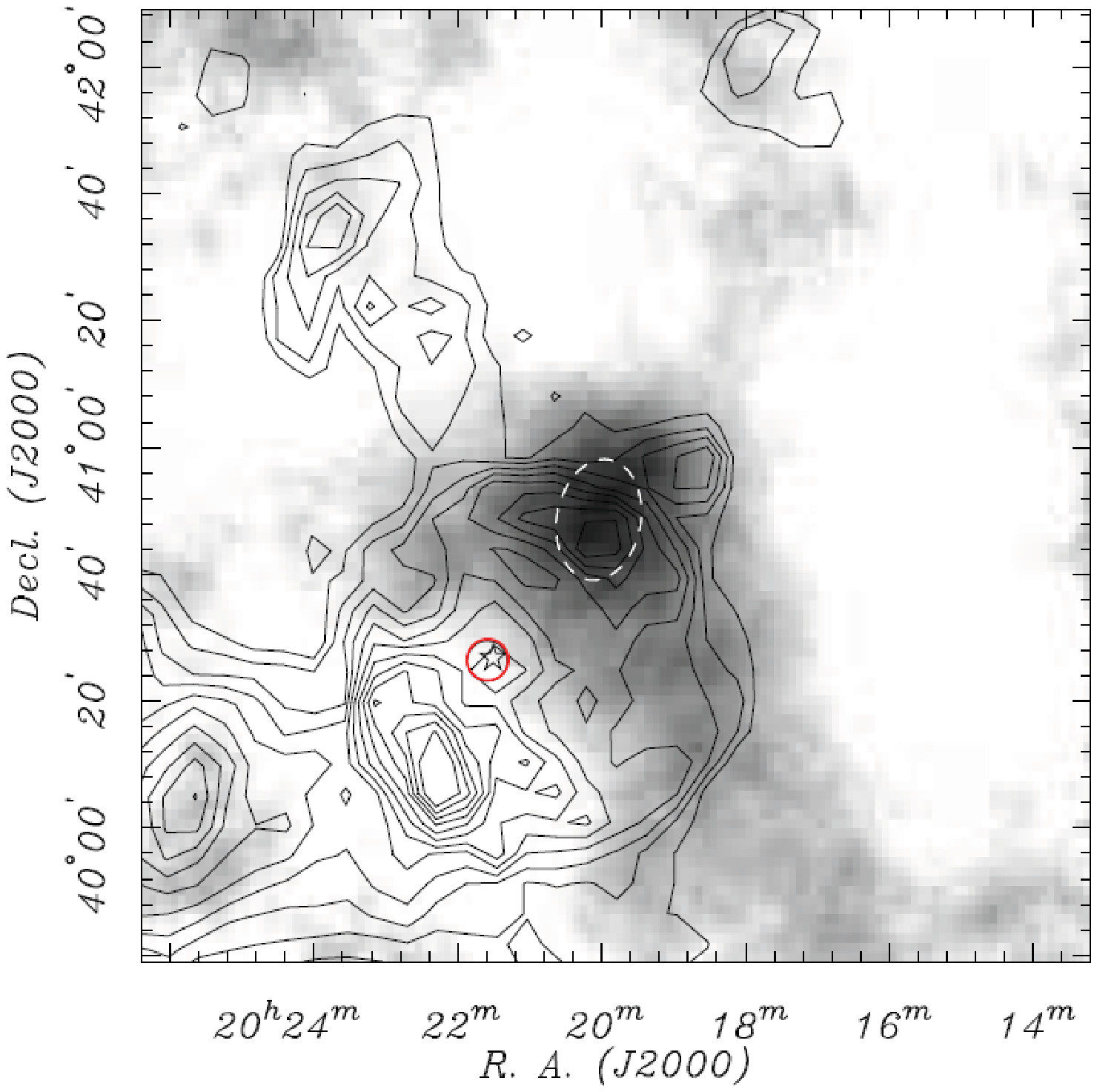}\label{verj2019_mw}}
  \caption{\subref{cygni_map} Significance map of the region around VER
  J2019+407, based on Fall 2009 observations alone.
  \subref{verj2019_mw}
  Acceptance-corrected excess map of the region around VER J2019+407, using all data taken through Fall 2009.  The
  black contours show the 1420MHz radio emission in the region (taken from the Canadian
  Galactic Plane Survey\cite{CGPS}) and clearly show the extent of the $\gamma$
  Cygni remnant.  The white-dashed line shows the $1\sigma$ ellipse from the
  VER J2019+407 source extension fit; the circle is the error circle of OFGL
  J2021.5+4026 and the star is the position of the associated Fermi pulsar.}
 \end{figure*}

The base survey had shown a hint of extended emission (at the
$4\sigma$ level) in the northwest corner of the $\gamma$ Cygni SNR;
survey-style follow-up data in Fall 2008 and wobble observations
taken in Spring 2009 raised the signal to the level of $\sim
5-7\sigma$ pre-trials in the various survey analyses.  In Fall 2009,
a set of independent wobble observations were taken at $0.6
^{\circ}$ offsets about the position (RA,Dec) $\rm
20^{h}19^{m}48^{s}$, $\rm Dec=40^{\circ}54'00''$.  The resulting
significance map is shown in Figure \ref{cygni_map}; extended
emission was detected with a maximum significance of 8.5 standard
deviation ($7.5\sigma$ post-trials) within a search region of radius
$0.25^{\circ}$ around the central wobble position, using the
standard extended source analysis from the survey.  The analysis and
search region were chosen \emph{a priori} based on the the previous
survey results.

The centroid and intrinsic extension of the source are based on the
Fall 2009 observations only, and were characterized by fitting an
asymmetric two-dimensional Gaussian convolved with the point-spread
function (PSF) of the instrument, to an acceptance-corrected
uncorrelated excess map with $0.05^{\circ}$ bins.  The PSF was
derived from a fit to Crab data taken in Fall 2009 and has a $68\%$
containment radius of $0.1^{\circ}$.  We find a preliminary centroid
position of $\rm RA=20^{h} 19^{m} 52.80^{s} \pm 0.017^{\circ} $,
$\rm Dec=40^{\circ} 47' 24.0'' \pm 0.023^{\circ}$ and a preliminary
extension of $0.16^{\circ} \pm 0.028^{\circ}$ and $0.11^{\circ} \pm
0.027^{\circ}$ along the major and minor axes.

The nature of VER J2019+407 is still unclear.  As the centroid of
VER J2019+407 is displaced from the Fermi pulsar position by over
$0.5^{\circ}$ (15pc for an assumed distance of 1.7kpc\cite{landp}),
an association between the two sources seems unlikely.  While it is
not uncommon for VHE gamma-ray pulsar wind nebulae to be displaced
from both the associated pulsar and the X-ray (or GeV gamma-ray)
counterpart of the PWN, a displacement of this size would be rare
and not easily justified. It seems more likely that the TeV emission
is in some way associated with the northwestern portion of the SNR
itself.  It would be attractive to interpret the emission as being
hadronic in origin and associated with a shock-cloud interaction,
but this would require sufficient density in HI to compensate for
the relative lack of molecular material, particularly CO, at the VER
J2019+407 position. There are two other puzzling features that must
also be explained: the reason for the correlation between the TeV
emission and the population of GeV electrons indicated by
high-intensity regions in the radio contours at 1420 MHz and 4.85
GHz,  and the confinement of emission to the northwestern portion of
the remnant. One possible scenario is that the SNR was previously
expanding within a low density bubble blown by winds from the
progenitor star. In this case, the SNR has only recently begun to
interact with relatively dense material in the interstellar medium,
with the northwestern portion of the remnant being the first to
encounter denser material.

Uchiyama \emph{et al.} have also argued for the presence of
shock-cloud interactions in the northwestern part of the remnant on
the basis of clumpy structures in 0.7-1keV ASCA satellite data, some
of which are co-located with VER J2019+407\cite{uchiyama}.  However,
the corresponding identification of hard X-ray sources in the same
region in the 4-7 keV band with shocked, high-density HI cloudlets
rested on the identification of this region with EGRET source 3EG
J2020+4017, which is now identified with OFGL J2021.5+4026 and
therefore now localized well away from VER J2019+407.  A Fermi upper
limit on GeV gamma-ray emission within the VER J2019+407 emission
region would therefore be essential to placing an upper limit on
possible HI cloudlet density in the region.

\section{Base Survey Sensitivity and Limits}

A set of detailed survey simulations, coupled with observations of
the Crab Nebula at multiple offsets, allows us to determine the
\emph{a priori} sensitivity of the survey analyses, not only to a
point source with a Crab-like spectrum, but to harder-spectrum
and/or significantly extended sources.  In order to best reproduce
the expected background conditions, blank (i.e. cosmic-ray
dominated) survey fields were used to provide the background for
these simulations.  Blank fields at an appropriate range of zenith
angles were pulled from the survey and arranged in a mocked-up
`cell' of the survey grid around a test point.  Showers that
reconstruct at a distance greater than $1.7^{\circ}$ from the center
of the field of view are not used in the analysis of survey data.
Therefore, only pointings where the center of the field of view is
less than $1.7^{\circ}$ from the test position are included in the
simulation, as only these pointings contribute significantly to the
sensitivity at the test point.


Simulated gamma rays were then injected into each field in the
simulated survey grid at the test position at a rate appropriate to
the source spectrum and flux strength in question.  In each case the
gamma rays were simulated at an appropriate camera offset and
matched as closely as possible to the background field in terms of
zenith angle and azimuth.  In the case of an extended source, the
injection positions were smeared by a two-dimensional Gaussian with
appropriate radii. The simulated grid is then analyzed using
standard survey analysis procedures.

 \begin{table}[!h]
  \caption{Survey Sensitivity}
  \label{sens_summary}
  \centering
  \begin{tabular}{|p{1.6cm}||p{1.6cm}|p{1.6cm}|p{1.6cm}|}
  \hline
    \multirow{2}{*}{Analysis} & \multicolumn{3}{|c|}{Source Properties}
    \\
     & Spectral Index & Extension (radius) & Flux   \\ \hline \hline
    Standard Point  & 2.5(2.0) & none & 4\% Crab $>$ 200
    GeV  \\
    Standard Extended & 2.5(2.0) & $0.2^{\circ}$ &  10\% Crab $>$ 200
    GeV \\ \hline
     Hard Point & 2.0 & none & 6.3\% Crab $>$ 500
    GeV \\
     Hard Extended & 2.0 & $0.2^{\circ}$  &  16\% Crab $>$ 500
    GeV
    \\ \hline

  \end{tabular}
  \end{table}

Preliminary results of the sensitivity studies, using a simulated
survey grid of four-telescope array observations taken at
$20^{\circ}$ zenith angle, have been completed.  Using the results
of theses studies, and accounting for variations in sensitivity due
to array configuration and zenith angle, we derive preliminary
values for the sensitivity of the various survey analyses to sources
of particular spectral indices and extensions, as summarized in
Table \ref{sens_summary}.
Extrapolating from these sensitivities and assuming a $3\sigma$
fluctuation, we calculate preliminary flux limits (for all points
below $3\sigma$ in the survey region) of $3\%$ of the Crab above 200
GeV for point sources and $8.5\%$ of the Crab above 200 GeV for
extended sources.

The base Cygnus region survey sees no points at or above $5\sigma$
pre-trials. If we extrapolate from our sensitivity estimates and the
population seen by the original H.E.S.S. survey\cite{hess_survey} of
the region between $-30 < l < 30$, and correct for the area covered
by the two surveys at or near full sensitivity, we would have
expected to see between two and three sources in the base survey.
While these preliminary results are not conclusive, they do seem to
indicate a difference in VHE gamma-ray source population between the
Cygnus region and the Galactic center region surveyed by H.E.S.S.

\section{Summary and Outlook}

VERITAS has completed a 140-hour survey of a 5 by 15 degree portion
of the Cygnus region of the Galactic Plane, and simulation studies
indicate that the base survey's average VHE point-source flux
sensitivity is better than $4\%$ of the Crab Nebula at energies
above 200 GeV, and we derive preliminary flux limits of $3\%$ of the
Crab Nebula for point sources and $8.5\%$ of the Crab Nebula for all
points in the survey below $3\sigma$.  Given our estimated
sensitivity and the known source population from the original 2005
H.E.S.S. survey of the Galactic Center region, we would have
expected to see 2-3 sources in the Cygnus region in the base survey
at or above $5\sigma$ pre-trials, but see none---a result indicative
of a population difference between the two regions.  Follow-up
observations within the survey region have also resulted in two
source detections: TeV J2032+4130 (which is likely related to the
coincident Fermi source OFGL J2032.2+4122) and a newly-discovered
region of extended emission, VER J019+407, overlapping the northwest
portion of the $\gamma$ Cygni SNR, which may or may not have a
relationship to the nearby Fermi source OFGL J2021.5+4026.

VERITAS observations continue on areas of interest within the Cygnus Region
survey.  Studies of correlations between VERITAS and Fermi maps in this region
are also planned.

\bigskip 
\begin{acknowledgments}

This research is supported by grants from the U.S. Department of
Energy, the US National Science Foundation, and the Smithsonian
Institution, by NSERC in Canada, by Science Foundation Ireland, and
by STFC in the UK.  We acknowledge the excellent work of the
technical support staff at the FLWO and the collaboration
institutions in the construction and operation of the instrument.

\end{acknowledgments}

\bigskip 

\end{document}